# Different Synthesis Routes of Graphene-based Metal Nanocomposites

Joydip Sengupta

Department of Electronic Science, Jogesh Chandra Chaudhuri College, Kolkata-700033

E-mail: joydipdhruba@gmail.com

Nanocomposite material proves to be the best candidate to match today's technological need as fascinating properties can be achieved by combining two or more nanomaterials. Among various nanomaterials, graphene is able to stand out far ahead of all others because of its novel structure and exclusive characteristics. In particular, graphene-metal nanocomposites have attracted enormous interest for their prospective use in various fields, including electronics, electrical and energy-related areas. However, for the utmost use of potential of graphene, it has to be homogenously embedded into metal matrices. Thus, appropriate synthesis route is decisive to obtain graphene-metal nanocomposites with desired properties. This chapter will summarize the different synthesis routes of high-quality graphene-metal nanocomposites along with their current developments.

## 1. Introduction

Development of new material is the fundamental culture of human race since its birth. Thus in order to accomplish unique properties, two or more materials can be combined resulting in a new material called composite material. Composite materials are solids that comprised of two or more different elements or phases, on a scale larger than the size of an atom. Since the birth of civilisation the synthesis of composite material started its journey with the invention of mud bricks and concrete. In general, composites are comprised of two types of materials, one is the called matrix or binder which encloses and binds together fragments or fibres of the other material, termed as reinforcement. If one of the constituent materials of the composite is of nanometer dimension then the composite is called nanocomposite[1] and in most of the cases the nanomaterial are used as reinforcement material. Since the advent of nanomaterial several of them is used to synthesise nanocomposite, as nanomaterial possesses remarkable improvement in their properties compared to their bulk counterpart[2]. The inclusion of nanomaterial as the reinforcement in the parent matrix also significantly enhances the properties of the resulting nanocomposite[3]. Among the different nanomaterials graphene has emerged as one of the most prospective nanomaterials due to the unique combination of its exotic properties[4]. Based on the binder material graphene nanocomposite can be categorized as graphene metal/metal oxide nanocomposite[5], graphene ceramic nanocomposite[6], graphene semiconductor nanocomposite[7] and graphene polymer nanocomposite[8]. Among various nanocomposites, graphene-metal/metal oxide composites have attracted considerable attention because of their potential applications in energy[9] and health[10]-related areas.

---

[1] Nakshatra Bahadur Singh and Sonal Agarwal, "Nanocomposites: An Overview," *Emerging Materials Research* 5 (2016): 5–43, doi:10.1680/jemmr.15.00025.

[2] Emil Roduner, "Size Matters: Why Nanomaterials Are Different," *Chemical Society Reviews* 35 (2006): 583–92, doi:10.1039/b502142c.

[3] Sanat K. Kumar and Ramanan Krishnamoorti, "Nanocomposites: Structure, Phase Behavior, and Properties," *Annual Review of Chemical and Biomolecular Engineering* 1, no. 1 (2010): 37–58, doi:10.1146/annurev-chembioeng-073009-100856.

[4] K. S. Novoselov et al., "A Roadmap for Graphene," *Nature* 490, no. 7419 (2012): 192–200, doi:10.1038/nature11458.

[5] Mujeeb Khan et al., "Graphene Based Metal and Metal Oxide Nanocomposites: Synthesis, Properties and Their Applications," *Journal of Materials Chemistry A* 3, no. 37 (2015): 18753–808, doi:10.1039/c5ta02240a.

[6] Kalaimani Markandan and Jit Kai, "Recent Progress in Graphene Based Ceramic Composites: A Review Kalaimani," *Journal of Materials Research* 32, no. 1 (2017): 84–106, doi:10.1557/jmr.2016.390.

[7] Ramin Yousefi and Mohsen Cheraghizade, *Semiconductor/Graphene Nanocomposites: Synthesis, Characterization, and Applications*, Applications of Nanomaterials (Elsevier Ltd., 2018), doi:10.1016/B978-0-08-101971-9.00002-8.

[8] Weifeng Chen et al., "A Critical Review on the Development and Performance of Polymer/graphene Nanocomposites," *IEEE Journal of Selected Topics in Quantum Electronics* 25, no. 6 (2018): 1059–73, doi:10.1515/secm-2017-0199.

[9] Yuezeng Su et al., "Two-Dimensional Carbon-Coated Graphene/metal Oxide Hybrids for Enhanced Lithium Storage," *ACS Nano* 6, no. 9 (2012): 8349–56, doi:10.1021/nn303091t.

[10] Md Azahar Ali et al., "Graphene Oxide-Metal Nanocomposites for Cancer Biomarker Detection," *RSC Advances* 7, no. 57 (2017): 35982–91, doi:10.1039/c7ra05491b.

However the quality of graphene and its homogeneous dispersion in the nanocomposite is the key to success.

## 2. Graphene and Its Synthesis

Graphene is an atomically thin 2D allotrope of carbon made of a single sheet of $sp^2$ hybridised carbon atoms, has exotic chemical and physical properties[11] with a planar density[12] of 0.77 mg/m$^2$. Graphene possesses strongest crystal structure[13], outstanding mechanical properties[14], extremely high thermal conductivity[15], electron mobility[16] and electrical conductivity[17].

---

[11] Andrea C. Ferrari et al., "Science and Technology Roadmap for Graphene, Related Two-Dimensional Crystals, and Hybrid Systems," *Nanoscale* 7, no. 11 (2015): 4598–4810, doi:10.1039/c4nr01600a.
[12] Q Li et al., "Highly Efficient Visible-Light-Driven Photocatalytic Hydrogen Production of CdS-Cluster-Decorated Graphene Nanosheets," *Journal of the American Chemical Society* 133, no. 28 (2011): 10878–84, doi:10.1021/ja2025454.
[13] G.H. Lee et al., "High-Strength Chemical-Vapor-Deposited Graphene and Grain Boundaries," *Science* 340, no. 6136 (2013): 1073–76, doi:10.1126/science.1235126.
[14] C. Lee et al., "Measurement of the Elastic Properties and Intrinsic Strength of Monolayer Graphene," *Science* 321, no. 5887 (2008): 385–88, doi:10.1126/science.1157996.
[15] S. Ghosh et al., "Extremely High Thermal Conductivity of Graphene: Prospects for Thermal Management Applications in Nanoelectronic Circuits," *Applied Physics Letters* 92, no. 15 (2008): 151911 (1-4), doi:10.1063/1.2907977.
[16] K. I. Bolotin et al., "Ultrahigh Electron Mobility in Suspended Graphene," *Solid State Communications* 146, no. 9–10 (2008): 351–55, doi:10.1016/j.ssc.2008.02.024.
[17] J. S. Choi et al., "Facile Fabrication of Properties-Controllable Graphene Sheet," *Scientific Reports* 6, no. December 2015 (2016): 1–7, doi:10.1038/srep24525.

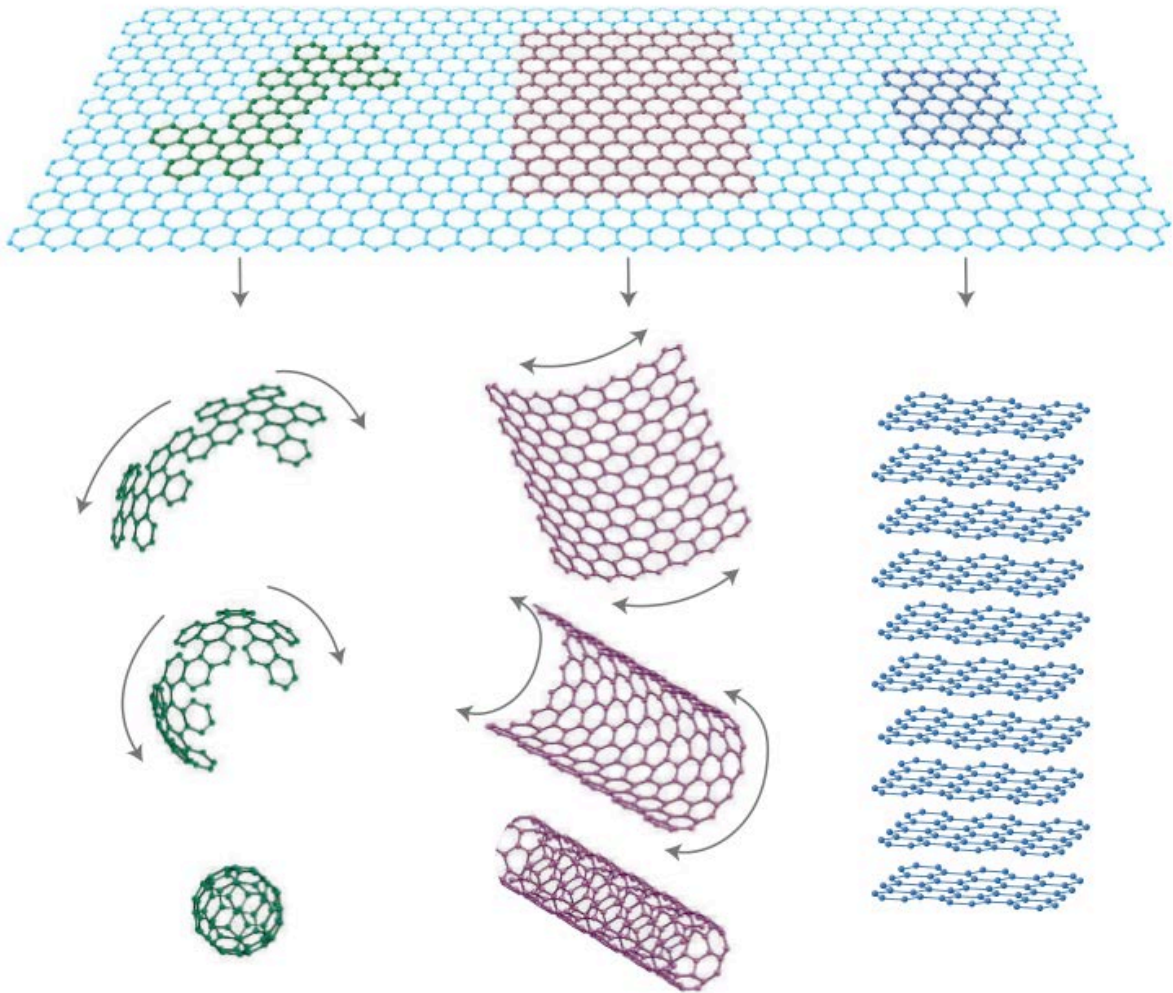

Fig. 1. Graphene is a 2D material that acts as a building block for carbonaceous materials of all other dimensions[18].

The graphene can be synthesised via top down processes like micromechanical exfoliation[19], electrochemical exfoliation[20], thermal exfoliation[21], reduction of graphene oxide[22], arc discharge[23], unzipping carbon nanotubes[24] and sonication[25]. The other graphene growth methods come under the category of bottom up process are pyrolysis[26], chemical vapor deposition[27] and epitaxial growth on silicon carbide[28]. There are good quality review papers[29,30] on the different synthesis processes of graphene and in particular the pros and cons of different synthesis routes are reported by Raccichini et al.[31].

---

[18] A. K. Geim & K. S. Novoselov, "The Rise of Graphene," *Nature Materials* 6 (2007): 183–91, doi:10.1038/nmat1849.
[19] K. S. Novoselov et al., "Electric Field Effect in Atomically Thin Carbon Films," *Science* 306, no. 5696 (2004): 666–69, doi:10.1126/science.1102896.
[20] A. M. Abdelkader et al., "How to Get between the Sheets: A Review of Recent Works on the Electrochemical Exfoliation of Graphene Materials from Bulk Graphite," *Nanoscale* 7, no. 16 (2015): 6944–56, doi:10.1039/C4NR06942K.
[21] X. V. Heerden, "The Influence of Three Different Intercalation Techniques on the Microstructure of Exfoliated Graphite" (University of Pretoria, Pretoria, South Africa, 2015).
[22] D.R. Dreyer et al., "The Chemistry of Graphene Oxide," *Chemical Society Reviews The* 39 (2010): 228–40, doi:10.1007/978-3-319-15500-5_3.

---

[23] K. S. Subrahmanyam et al., "Simple Method of Preparing Graphene Flakes by an Arc-Discharge Method," *J. Phys. Chem. C* 113, no. February (2009): 4257–59, doi:10.1021/jp900791y.
[24] P. Kumar et al., "Laser-Induced Unzipping of Carbon Nanotubes to Yield Graphene Nanoribbons," *Nanoscale* 3, no. 5 (2011): 2127–2129, doi:10.1039/c1nr10137d.
[25] C. Shih et al., "Understanding the Stabilization of Liquid-Phase Exfoliated Graphene in Polar Solvents: Molecular Dynamics Simulation and Kinetic Theory of Colloid Aggregation," *Journal of the American Chemical Society* 132 (2010): 14638–14648, doi:10.1021/nl903557p.
[26] Juan Yang et al., "Efficient Synthesis of Graphene-Based Powder via in Situ Spray Pyrolysis and Its Application in Lithium Ion Batteries," *RSC Advances* 3, no. 37 (2013): 16449–55, doi:10.1039/c3ra41724g.
[27] M. Losurdo et al., "Graphene CVD Growth on Copper and Nickel: Role of Hydrogen in Kinetics and Structure," *Physical Chemistry Chemical Physics* 13, no. 46 (2011): 20836–43, doi:10.1039/c1cp22347j.
[28] M. Suemitsu et al., "Epitaxial Graphene Formation on 3C-SiC/Si Thin Films," *Journal of Physics D: Applied Physics* 47, no. 9 (2014): 094016 (1-11), doi:10.1088/0022-3727/47/9/094016.
[29] H. Cheun Lee et al., "Review of the Synthesis, Transfer, Characterization and Growth Mechanisms of Single and Multilayer Graphene," *RSC Advances* (Royal Society of Chemistry, 2017), doi:10.1039/C7RA00392G.
[30] Md. Sajibul Alam Bhuyan et al., "Synthesis of Graphene," *Nternational Nano Letters* 6 (2016): 65–83, doi:10.1007/s40089-015-0176-1.
[31] Rinaldo Raccichini et al., "The Role of Graphene for Electrochemical Energy Storage," *Nature Materials* 14, no. 3 (2015): 271–79, doi:10.1038/nmat4170.

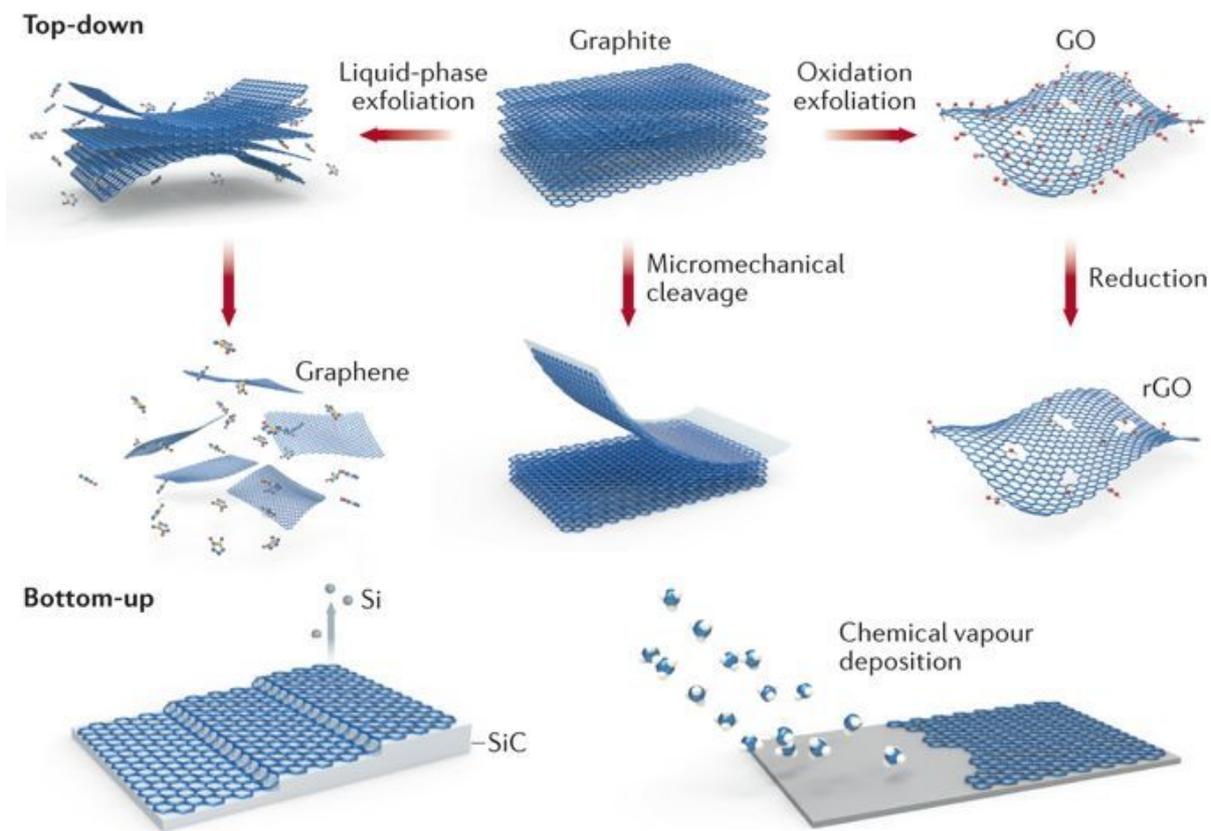

Fig. 2. Schematic representation of the graphene synthesis methods[32].

## 3. Synthesis of graphene-metal/metal oxide nanocomposites

There are various routs to synthesize graphene-metal/metal oxide nanocomposites as shown in the illustration below.

---

[32] Xiao-Ye Wang, Akimitsu Narita, and Klaus Müllen, "Precision Synthesis versus Bulk-Scale Fabrication of Graphenes," *Nature Reviews Chemistry* 2, no. December (2017): 0100 (1-10), doi:10.1038/s41570-017-0100.

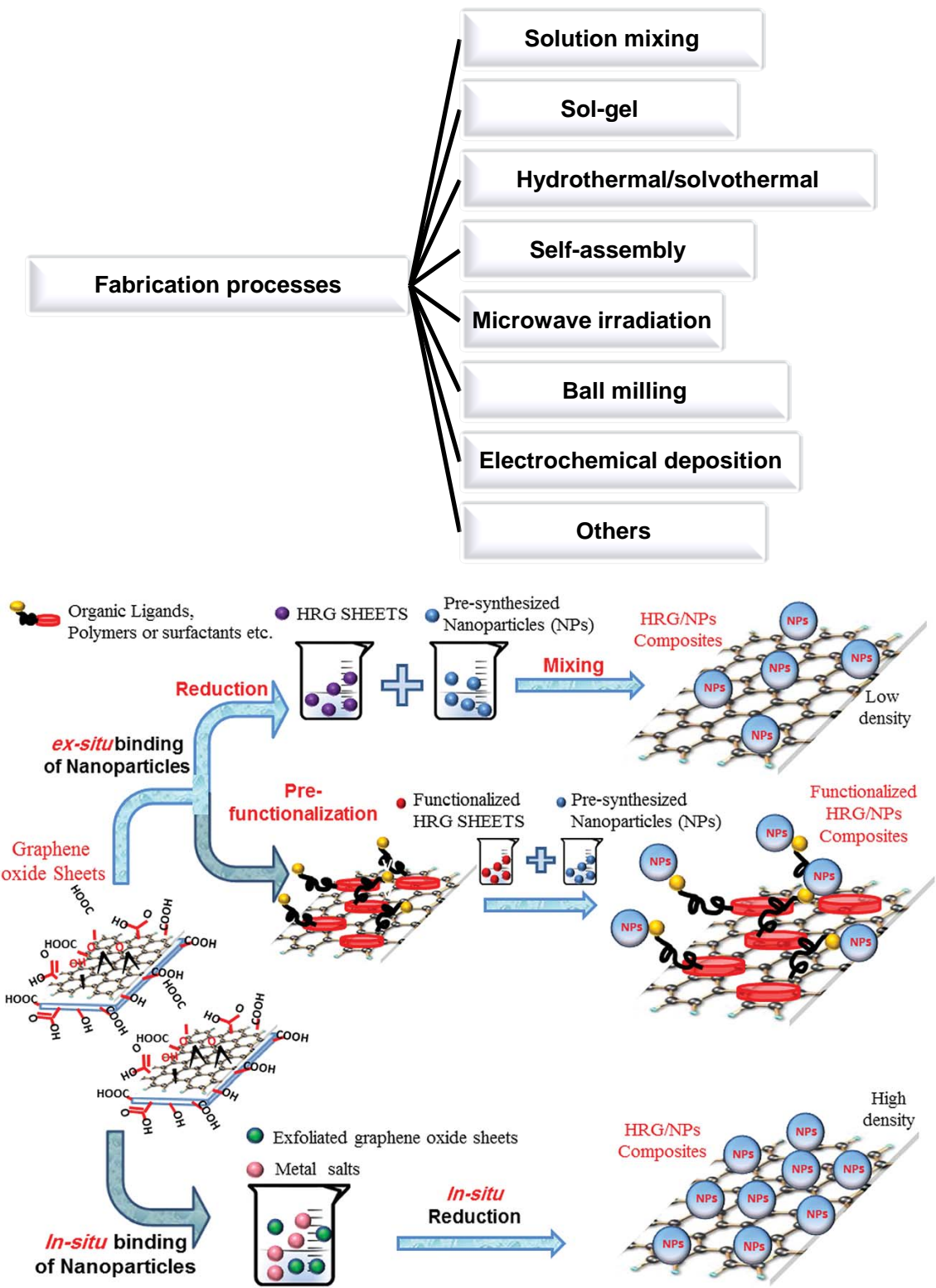

Fig. 3. Schematic illustration of binding mechanisms of the nanoparticles onto rGO sheets[33].

---

[33] Khan et al., "Graphene Based Metal and Metal Oxide Nanocomposites: Synthesis, Properties and Their Applications."

## 3.1 Solution Mixing Method

Solution mixing method is widely used to synthesise of graphene-metal/metal oxide nanocomposite as it can be performed at low temperature, promotes faster de-aggregation and dispersion of reinforcement material and produce composites with uniform reinforcement dispersion[34]. Tang et al.[35] used in situ chemical reduction method to prepare graphene nanosheets (GNS) decorated with Ni nanoparticles and afterwards GNS-Ni hybrids were wet-mixed with electrolytic Cu powder to produce GNS–Ni/Cu nanocomposite. The mechanical properties of the resulting nanocomposite revealed high Young's modulus of 132 GPa and yield strength of 268 MPa. Li et al.[36] used natural graphite powder to prepare graphene oxide (GO) via Hummer's method[37] and synthesised Ni nanoparticles decorated graphene platelets (GPL) using chemical reduction of Ni ions on the surface of the GO. Afterwards Cu powder was solution mixed with Ni nanoparticles decorated GPL to synthesise Ni-GPL-Cu nanocomposite with ultimate tensile strength of 250 MPa. Wang et al.[38] used reduced graphene oxide (rGO) and $Cu(OH)_2$ composite sheets to fabricate micro-layered structure of rGO-Cu powder and sintered it with spark plasma. The tensile strength of resulting nanocomposite was 608 MPa which is more than three times higher than that of the Cu matrix. Zeng et al.[39] added Al powder in the aqueous GO suspension which was followed by ultrasonic treatment, stirring and drying to obtain graphene-Al nanocomposites with ultimate tensile strength of 255 MPa. Other than improvement in mechanical properties graphene metal/metal oxide nanocomposites are also used to address different environmental and energy related issues. Peak et al.[40] used ethylene glycol to disperse chemically reduced GNS and later on reassembled it in the presence of $SnO_2$ nanoparticles to synthesise graphene-$SnO_2$ nanocomposite with superior cyclic performances. Williams et al[41] proposed an efficient way to fabricate graphene-$TiO_2$ nanocomposites via UV-assisted photocatalytic reduction of GO in solution phase. Bell et al.[42] used a solution phase photocatalytic reduction method to integrate rGO into $TiO_2$ film to manufacture photovoltaic cells that exhibited 10-fold increase in the photo-generated current that of $TiO_2$ only. Akhavan et al.[43] used the similar methodology for the formation of GO-$TiO_2$ thin film and studied the antibacterial activity of the thin films against the Escherichia coli bacteria.

---

[34] S. Mohan et al., "Rheology and Processing of Inorganic Nanomaterials and Quantum Dots/Polymer Nanocomposites," *Rheology and Processing of Polymer Nanocomposites*, 2016, 355–82, doi:10.1002/9781118969809.ch10.

[35] Yanxia Tang et al., "Enhancement of the Mechanical Properties of Graphene-Copper Composites with Graphene-Nickel Hybrids," *Materials Science and Engineering A* 599 (2014): 247–54, doi:10.1016/j.msea.2014.01.061.

[36] Meixia Li et al., "Highly Enhanced Mechanical Properties in Cu Matrix Composites Reinforced with Graphene Decorated Metallic Nanoparticles," *Journal of Materials Science* 49, no. 10 (2014): 3725–31, doi:10.1007/s10853-014-8082-x.

[37] W. S. Hummers et al., "Preparation of Graphitic Oxide," *Journal of the American Chemical Society* 80, no. 6 (1958): 1339, doi:10.1021/ja01539a017.

[38] Lidong Wang et al., "Graphene-Copper Composite with Micro-Layered Grains and Ultrahigh Strength," *Scientific Reports* 7 (2017): 41896 (1-10), doi:10.1038/srep41896.

[39] Xiang Zeng et al., "Fabrication of Homogeneously Dispersed graphene/Al Composites by Solution Mixing and Powder Metallurgy," *International Journal of Minerals, Metallurgy and Materials* 25, no. 1 (2018): 102–9, doi:10.1007/s12613-018-1552-4.

**3.2 Sol-gel method**

The sol–gel route[44] provides simple, inexpensive preparation of a homogeneous composite material with excellent compositional control and thus used in many applications[45]. Zhang et al.[46] used tetrabutyl titanate and GO as the starting materials for the sol gel synthesis of graphene-$TiO_2$ nanocomposites and found that the photocatalytic activity of the composite was affected by both graphene content and the calcinations atmosphere. Li et al.[47] used sol gel method to ultra-disperse $TiO_2$ nanoparticles on graphene with extreme control so that the grown material can achieve twice the specific capacity as that of mechanically mixed composites. Innocenzi et al.[48] made optically transparent graphene-$SiO_2$ nanocomposite via sol gel process using 1-vinyl-2-pyrrolidone and $SiCl_4$ to avoid aggregation of graphene sheet. The grown nanocomposites can be potentially used in optical limiters. Hintze et al.[49] fabricated controlled carbon phase based rGO-$SiO_2$ nanocomposite using sol-gel technique with appreciable electrical conductivity. Rezaei et al.[50] used sol-gel process to fabricate graphene-$SiO_2$ nanocomposite with remarkable specific capacitance of 428 $Fg^{-1}$ at current density of $1Ag^{-1}$ and good cycling stability. Graphene-TiC nanocomposites was synthesised by Wang et al.[51] for shock/impact absorption via sol gel process using Furfurly alcohol (FA) as a carbon source. A nanocomposite of functional GNS and cobalt sulphide with high reversible capacity of 466 $mAhg^{-1}$ and good rate capability was prepared by Chen et al.[52]. Damavandi et al[53] used graphene and carbon nanotube (CNT) as the reinforcement material in $Al_2O_3$ matrix and the resulted sol- gel derived nanocomposite exhibited fracture toughness and flexural strength of 6.2 $MPam^{1/2}$ and 420 MPa, respectively. Graphene-$BiFO_3$ composite was prepared by Nayak et al.[54] via sol gel process for electrochemical supercapacitor application as it revealed specific capacitance of 17-4 mF/cm$^2$ and 95% retention of capacitance after 2000 cycles. Giampiccolo et al.[55] synthesised graphene-TiO$_2$ nanocomposite via sol-gel route for the sensing of harmful gas NO$_2$ with the detection limit of about 50 ppb to provide a safer environment.

---

[40] Seung-min Paek, Eunjoo Yoo, and Itaru Honma, "Enhanced Cyclic Performance and Lithium Storage Capacity of SnO2/ Graphene Nanoporous Electrodes with," *Nano Lett.* 9, no. 1 (2009): 72–75, doi:10.1021/nl802484w.
[41] Graeme Williams, "UV-Assisted Photocatalytic Reduction of Graphene Oxide," *ACS Nano* 2, no. 7 (2008): 1487–91, doi:10.1021/nn800251f.
[42] Nicholas J. Bell et al., "Understanding the Enhancement in Photoelectrochemical Properties of Photocatalytically Prepared TiO2-Reduced Graphene Oxide Composite," *Journal of Physical Chemistry C* 115, no. 13 (2011): 6004–9, doi:10.1021/jp1113575.
[43] O Akhavan and E Ghaderi, "Photocatalytic Reduction of Graphene Oxide Nanosheets on TiO2 Thin Film for Photoinactivation of Bacteria in Solar Light Irradiation," *J. Phys. Chem. C* 113 (2009): 20214–20, doi:10.1021/jp906325q.
[44] Larry L. Hench and Jon K. West, "The Sol-Gel Process," *Chemical Reviews* 90, no. 1 (1990): 33–72, doi:10.1021/cr00099a003.
[45] Amir Dehghanghadikolaei, Jamal Ansary, and Reza Ghoreishi, "Sol-Gel Process Applications: A Mini-Review," *Proceedings of the Nature Research Society* 2, no. June (2018): 02008 (1-11), doi:10.11605/j.pnrs.201802008.
[46] Xiao Yan Zhang et al., "Graphene/$TiO_2$ nanocomposites: Synthesis, Characterization and Application in Hydrogen Evolution from Water Photocatalytic Splitting," *Journal of Materials Chemistry* 20, no. 14 (2010): 2801–6, doi:10.1039/b917240h.

---

[47] Wei Li et al., "Sol-Gel Design Strategy Towards Ultra-Dispersed TiO2 Nanoparticles on Graphene for High-Performance Lithium Ion Batteries," *Journal of the American Chemical Society* 135, no. 49 (2013): 18300–18303, doi:10.1021/ja4100723.
[48] Plinio Innocenzi et al., "Sol-Gel Chemistry for Graphene-Silica Nanocomposite Films," *New Journal of Chemistry* 38, no. 8 (2014): 3777–82, doi:10.1039/c4nj00535j.
[49] Cornelia Hintze et al., "Facile Sol-Gel Synthesis of Reduced Graphene Oxide/silica Nanocomposites," *Journal of the European Ceramic Society* 36, no. 12 (2016): 2923–30, doi:10.1016/j.jeurceramsoc.2015.11.033.
[50] Rahim Mohammad-Rezaei and Habib Razmi, "Preparation and Characterization of Reduced Graphene Oxide Doped in Sol-Gel Derived Silica for Application in Electrochemical Double-Layer Capacitors," *International Journal of Nanoscience and Nanotechnology* 12, no. 4 (2016): 233–41.
[51] Xiaojing Wang et al., "Graphene/titanium Carbide Composites Prepared by Sol–gel Infiltration and Spark Plasma Sintering," *Ceramics International* 42, no. 1 (2016): 122–31, doi:10.1016/j.ceramint.2015.08.017.
[52] Tingting Chen et al., "A Facile Sol-Gel Route to Prepare Functional Graphene Nanosheets Anchored with Homogeneous Cobalt Sulfide Nanoparticles as Superb Sodium-Ion Anodes," *Journal of Materials Chemistry A* 5, no. 7 (2017): 3179–85, doi:10.1039/C6TA10272G.
[53] Yazdani B Damavandi et al., "A Sol-Gel Approach to Produce Highly Dispersed Graphene Oxide-Carbon Nanotube Hybrid Reinforcement into Alumina Matrix," *VERUSCRIPT FUNCTIONAL NANOMATERIALS*, 2017, 1–17, doi:10.22261/BKE2BP.

**3.3 Hydrothermal/ Solvothermal method**

Hydrothermal and solvothermal synthesis are the methods of producing composites generally in an autoclave under high temperatures and pressures[56]. Aqueous solution is used hydrothermal method while in solvothermal method non-aqueous solution is used. GO-TiO$_2$ nanocomposite was synthesised via two step hydrothermal method by Liang et al.[57] which showed superior photocatalytic activity to other forms of TiO$_2$. Yang et al.[58] deposited Ag nanoparticle on graphene using solvothermal method to fabricate graphene-Ag nanocomposites with excellent electroconductibility using ethylene glycol or deionzed water/hydrazine. The hydrazine proved to be better reducing agent thus had more control over the size and morphology of Ag nanoparticle. Lin-jun et al.[59] did similar experiment and the resulted graphene-Ag nanocomposites showed electroconductibility of 2.94scm$^{-1}$. A novel

---

[54] S. Nayak et al., "Sol–gel Synthesized BiFeO3–Graphene Nanocomposite as Efficient Electrode for Supercapacitor Application," *Journal of Materials Science: Materials in Electronics* 29, no. 11 (2018): 9361–68, doi:10.1007/s10854-018-8967-6.

[55] Andrea Giampiccolo et al., "Sol Gel graphene/TiO2 Nanoparticles for the Photocatalytic-Assisted Sensing and Abatement of NO2," *Applied Catalysis B: Environmental* 243, no. 2 (2019): 183–94, doi:10.1016/j.apcatb.2018.10.032.

[56] S. H. Feng and G. H. Li, *Hydrothermal and Solvothermal Syntheses*, *Modern Inorganic Synthetic Chemistry: Second Edition*, 2017, doi:10.1016/B978-0-444-63591-4.00004-5.

[57] Yongye Liang et al., "TiO 2 Nanocrystals Grown on Graphene as Advanced Photocatalytic Hybrid Materials," *Nano Research* 3, no. 10 (2010): 701–5, doi:10.1007/s12274-010-0033-5.

[58] Juan Yang et al., "Synthesis of graphene/Ag Nanocomposite with Good Dispersibility and Electroconductibility via Solvothermal Method," *Materials Chemistry and Physics* 129, no. 1–2 (2011): 270–74, doi:10.1016/j.matchemphys.2011.04.002.

[59] Huang Lin-jun et al., "Synthesis of Graphene/Metal Nanocomposite Film With Good Dispersibility via Solvothermal Method," *International Journal of ELECTROCHEMICAL SCIENCE* 7 (2012): 11068–75.

hydrothermal route was devised by Lee et al.[60] to synthesise graphene wrapped $TiO_2$ nanoparticle for enhancing the photocatalytic activity of $TiO_2$ under visible light. For precise detection of nonenzymatic glucose Song et al.[61] synthesised GO-CuO nanocomposite using hydrothermal method and also varied the different parameters such as hydrothermal temperature, amount and structure of CuO in order to optimise the process and the grown material showed the possibility to be used for highly enhanced nonenzymatic glucose detection. Graphene-$CeO_2$ nanocomposite with high catalytic activity was prepared by Srivastava et al.[62] using hydrothermal method. Zhang et al.[63] synthesized rGO-$Fe_3O_4$ nanocomposite using one pot hydrothermal route which depicted high capacity of 884 $mAhg^1$ after 100 cycles indicating its potential application as anode material of LIBs. Graphene-$Mn_3O_4$ nanocomposite was synthesised by Liu et al.[64] via solvothermal process in ethanol solution. The grown material is a promising candidate for supercapacitor as with the mass percent of $Mn^{2+}$:GO (10:90) it exhibited high specific capacitance (~245 F/g) at 5mV/s. Ma et al.[65] developed an unique solvothermal synthesis route to prepare self-assembled three dimensional graphene-CoO nanocomposite with enhanced electrochemical performance for use in lithium ion battery (LIB). Graphene-$MnFe_2O_4$ nanocomposite was prepared by Chella

---

[60] Joon Seok Lee, Kyeong Hwan You, and Chan Beum Park, "Highly Photoactive, Low Bandgap TiO2nanoparticles Wrapped by Graphene," *Advanced Materials* 24, no. 8 (2012): 1084–88, doi:10.1002/adma.201104110.

[61] Jian Song et al., "Synthesis of Graphene Oxide Based Cuo Nanoparticles Composite Electrode for Highly Enhanced Nonenzymatic Glucose Detection," *ACS Applied Materials and Interfaces* 5, no. 24 (2013): 12928–34, doi:10.1021/am403508f.

[62] Manish Srivastava et al., "A Facile One-Step Hydrothermal Synthesis of Graphene/CeO2 Nanocomposite and Its Catalytic Properties," *Advanced Materials Research* 747 (2013): 242–45, doi:10.4028/www.scientific.net/AMR.747.242.

[63] Xiao Zhang et al., "One-Pot Hydrothermal Synthesis of Fe3O4/reduced Graphene Oxide Nanocomposite for Enhanced Lithium Storage," *Indian Journal of Chemistry - Section A Inorganic, Physical, Theoretical and Analytical Chemistry* 53, no. 3 (2014): 265–73.

[64] Y. F. Liu et al., "Solvothermal Synthesis of Mn3O4nanoparticle/graphene Sheet Composites and Their Supercapacitive Properties," *Journal of Nanomaterials*, 2014, 190529 (1-11), doi:10.1155/2014/190529.

[65] Jingjing Ma et al., "A Solvothermal Strategy: One-Step in Situ Synthesis of Self-Assembled 3D Graphene-Based Composites with Enhanced Lithium Storage Capacity," *Journal of Materials Chemistry A* 2, no. 24 (2014): 9200–9207, doi:10.1039/c4ta01006j.

et al.[66] using solvothermal method and investigated its adsorbent and antimicrobial properties. The examination revealed that prepared nanocomposite had good adsorption efficiency regarding the removal of toxic heavy metal ions (Pb and Cd) and can also act as an efficient antimicrobial agent. Zhang et al.[67] synthesized graphene- $CuFe_2O_4$ nanocomposite employing solvothermal process and studied the electrochemical properties and the material exhibited electrochemical capacitance of 576.6 F·g$^{-1}$ at current density 1 A·g$^{-1}$ with high rate performance and cycling stability. Graphene-NiO nanocomposite was prepared by Narasimharao et al.[68] via hydrothermal synthesis which was of high potential for application in supercapacitor as electrode. Venkateshalu et al.[69] synthesised rGO-$CoS_2$ nanocomposite via hydrothermal process for application in supercapacitor and the grown material showed a specific capacitance of 28 F/g at a current density of 0.5 A/g. N-doped graphene-$Fe_2O_3$ composites was synthesised via hydrothermal process by Pu et al.[70] using melamine as the nitrogen source. The resultant material showed a remarkable specific capacitance values at low and high current densities 698 (1 A/g) and 354 (20 A/g) F/g, respectively and established its candidature as material for supercapacitor. Sagadevan et al.[71] prepared graphene-$SnO_2$ nanocomposite using hydrothermal approach with high photocatalytic activity and the synthesised product can be potentially used for industrial waste water management. GO-

---

[66] Santhosh Chella et al., "Solvothermal Synthesis of MnFe2O4-Graphene Composite-Investigation of Its Adsorption and Antimicrobial Properties," *Applied Surface Science* 327 (2015): 27–36, doi:10.1016/j.apsusc.2014.11.096.

[67] Wang Zhang et al., "One-Step Facile Solvothermal Synthesis of Copper Ferrite-Graphene Composite as a High-Performance Supercapacitor Material," *ACS Applied Materials and Interfaces* 7, no. 4 (2015): 2404–14, doi:10.1021/am507014w.

[68] Narasimharao K et al., "Synthesis of Graphene Oxide by Modified Hummers Method and Hydrothermal Synthesis of Graphene-NiO Nano Composite for Supercapacitor Application," *Journal of Material Science & Engineering* 5, no. 6 (2016): 1000284, doi:10.4172/2169-0022.1000284.

[69] Sandhya Venkateshalu, Dinesh Rangappa, and Andrews Nirmala Grace, "Hydrothermal Synthesis and Electrochemical Properties of CoS2 - Reduced Graphene Oxide Nanocomposite for Supercapacitor Application," *International Journal of Nanoscience* 16, no. 3 (2017): 1760020 (1-8), doi:10.1142/S0219581X17600201.

[70] Nen Wen Pu et al., "Hydrothermal Synthesis of N-Doped graphene/Fe2O3nanocomposite for Supercapacitors," *International Journal of Electrochemical Science* 13, no. 7 (2018): 6812–23, doi:10.20964/2018.07.16.

[71] Suresh Sagadevan et al., "A Facile Hydrothermal Approach for Catalytic and Optical Behavior of Tin Oxide-Graphene (SnO2/G) Nanocomposite," *PLoS ONE* 13, no. 10 (2018): 1–15, doi:10.1371/journal.pone.0202694.

$MnO_2$ nanocomposite was synthesised by Liang et al.[72] via one step hydrothermal process, which can used as adsorbent for the removal of heavy metal ions ($Pb^{2+}$, $Cu^{2+}$, $Cd^{2+}$ and $Zn^{2+}$) from aqueous solution.

**3.4 Self assembly method**

Self assembly[73] is a process in which a system's constituent reorganizes into some specific ordered or structure or pattern as a result of precise local interactions among the constituents themselves. Wang et al.[74] used anionic sulfate surfactants to stabilize graphene in aqueous medium and finally to felicitate self assembled growth of graphene-$TiO_2$ nanocomposite which can be potentially used in LIB for its high specific capacity. Liu et al.[75] devise an approach to self-assemble $TiO_2$ nanorods on GO sheets and the resulted nanocomposite showed enhanced photocatalytic activity under UV irradiation. Yang et al.[76] devised a novel strategy for the formation of graphene-metal oxide nanoparticles ($SiO_2$ and $Co_3O_4$) core-shell hybrids by electrostatic forces leading to notable lithium-storage performance, inclusive of high reversible capacity and exceptional cycle performance. Wang et al.[77] demonstrated a ternary self-assembly route employing graphene as basic building blocks to synthesise ordered graphene- metal oxide nanocomposite. One of such kind of material, graphene-$SnO_2$ nanocomposite exhibited near theoretical specific energy density for lithium ion insertion/extraction without significant degradation of charge/discharge. Li et al.[78] reported a facile one-pot, template-free self-assembly production route to synthesise graphene-$TiO_2$ nanocomposite. The nanocomposite depicted considerable enhancement in lithium specific capacity, high photocatalytic activity in removing organic pollutant and finally hydrogen evolution by splitting the water. Du et al.[79] fabricated hierarchically ordered macro-mesoporous graphene-$TiO_2$ nanocomposite via self assembly which showed the characteristic of rapid adsorbation and photodegradation of organic dyes for potential application in waste water treatment. Electrostatically derived self-assembled growth of titanate and rGO was carried out by Kim et al.[80] which revealed an enhanced performance in the photodegradation of organic molecules. Guoxin et al.[81] prepared rGO- $Fe_2O_3$ nanocomposites via self assembly method in aqueous phase through electrostatic attraction. Lin et al.[82] synthesised graphene-Co nanocomposite via carbonization a self-assembly constituted of a Co-based metal organic framework, ZIF-67, and GO. The grown material can serve as the effective catalyst to activate peroxymonosulfate in the advanced oxidation process. Xu et al.[83] prepared self assembled Ag nanowires-graphene nanocomposites for use in electrically conductive adhesives to improve its electrical conductivity and mechanical properties. Zhang et al.[84]

---

[72] Chunyan Liang et al., "Facile One-Step Hydrothermal Syntheses of Graphene oxide–MnO2 Composite and Their Application in Removing Heavy Metal Ions," *Micro & Nano Letters* 13, no. 8 (2018): 1179–84, doi:10.1049/mnl.2017.0761.
[73] Anne-Caroline Genix, "Nanoparticle Self-Assembly: From Interactions in Suspension to Polymer Nanocomposites," *Soft Matter* 14 (2018): 5161–79, doi:10.1039/C8SM00430G.
[74] Donghai Wang et al., "Self-Assembled TiO2–Graphene Hybrid Nanostructures for Enhanced Li-Ion Self-Assembled TiO2–Graphene Hybrid Insertion," *ACS Nano* 3, no. 4 (2009): 907–14, doi:10.1021/nn900150y.
[75] Jincheng Liu et al., "Self-Assembling TiO2nanorods on Large Graphene Oxide Sheets at a Two-Phase Interface and Their Anti-Recombination in Photocatalytic Applications," *Advanced Functional Materials* 20, no. 23 (2010): 4175–81, doi:10.1002/adfm.201001391.
[76] Shubin Yang et al., "Fabrication of Graphene-Encapsulated Oxide Nanoparticles: Towards High-Performance Anode Materials for Lithium Storage," *Angewandte Chemie - International Edition* 49, no. 45 (2010): 8408–11, doi:10.1002/anie.201003485.
[77] Donghai Wang et al., "Ternary Self-Assembly of Ordered Metal Oxide-Graphene Nanocomposites for Electrochemical Energy Storage," *ACS Nano* 4, no. 3 (2010): 1587–95, doi:10.1021/nn901819n.

---

[78] Na Li et al., "Battery Performance and Photocatalytic Activity of Mesoporous Anatase TiO2nanospheres/graphene Composites by Template-Free Self-Assembly," *Advanced Functional Materials* 21, no. 9 (2011): 1717–22, doi:10.1002/adfm.201002295.

[79] Jiang Du et al., "Hierarchically Ordered Macro- Mesoporous TiO2-Graphene Composite Films: Improved Mass Transfer, Reduced Charge Recombination, and Their Enhanced Photocatalytic Activities," *ACS Nano* 5, no. 1 (2011): 590–96, doi:10.1021/nn102767d.

[80] In Young Kim et al., "A Strong Electronic Coupling between Graphene Nanosheets and Layered Titanate Nanoplates: A Soft-Chemical Route to Highly Porous Nanocomposites with Improved Photocatalytic Activity," *Small* 8, no. 7 (2012): 1038–48, doi:10.1002/smll.201101703.

[81] Hu Guoxin and Zhengxia Xu, "Monodisperse Iron Oxide Nanoparticle-Reduced Graphene Oxide Composites Formed by Self-Assembly in Aqueous Phase," *Fullerenes Nanotubes and Carbon Nanostructures* 23, no. 4 (2014): 283–89, doi:10.1080/1536383X.2013.812633.

[82] Kun Yi Andrew Lin, Fu Kong Hsu, and Wei Der Lee, "Magnetic Cobalt-Graphene Nanocomposite Derived from Self-Assembly of MOFs with Graphene Oxide as an Activator for Peroxymonosulfate," *Journal of Materials Chemistry A* 3, no. 18 (2015): 9480–90, doi:10.1039/c4ta06516f.

[83] Tao Xu et al., "Self-Assembly Synthesis of Silver Nanowires/ Graphene Nanocomposite and Its Effects on the Performance of Electrically Conductive Adhesive," *Materials* 11, no. 10 (2018): 2028 (1-16), doi:10.3390/ma11102028.

[84] Qi Zhang et al., "Self-Assembly of Graphene-Encapsulated Cu Composites for Nonenzymatic Glucose Sensing," *ACS Omega* 3 (2018): 3420–28, doi:10.1021/acsomega.7b01197.

prepared rGO-encapsulated Cu nanoparticles employing electrostatic self assembly procedure which revealed exceptional electrocatalytic activity towards glucose oxidation along with a wide detection range, low detection limit and high sensitivity.

**3.5 Microwave irradiation method**

Microwave radiation[85] generates rapid intense heating in the interior of the sample and consequently reduces reaction times to provide cleaner reactions environment with energy saving. Zhang et al.[86] prepared graphene-magnetite nanocomposites employing microwave radiation which depicted high reversible as well as rate capacities (350 mAhg$^{-1}$) and enhanced cycling performances (650 mAhg$^{-1}$) thus suitable to be used as anode material of LIBs. Graphene-MnO$_2$ nanocomposite was synthesized using microwave irradiation by Yan et al.[87] which showed enhanced high rate electrochemical performance with specific capacitance as high as 310 Fg$^{-1}$. The same group[88] also prepared graphene-Co$_3$O$_4$ nanocomposite having long cycle life via microwave irradiation, however the obtained maximum specific capacitance was 243.2 Fg$^{-1}$. Graphene-metal (Ag, Au, Pt and Pd) nanocomposites was synthesized by Subrahmanyam et al.[89] under microwave irradiation which displayed significant electronic interaction between the constituents resulted from the ionization energy and electron affinity of the metal particles. Marquardt et al.[90] used stable Ru or Rh metal nanoparticles to produce graphene metal nanocomposite via microwave irradiation which were capable to act as catalyst in hydrogenation reactions resulting in complete conversion of cyclohexene or benzene to cyclohexane under organic-solvent-free conditions. A simple two-step microwave assisted synthesis of GO-$Fe_2O_3$ nanocomposite was carried out by Zhu et al.[91] which displayed discharge and charge capacities of 1693 and 1227 mAh/g, respectively with enhanced cycling performance and rate capability indicating its potential use as electrode materials for LIBs. Hsu et al.[92] synthesized rGO-Ag nanocomposite via microwave irradiation for use as surface-enhanced Raman scattering (SERS) substrate with high homogeneity and sensitivity. Three dimensional expanded graphene-metal oxides ($Fe_2O_3$ and $MnO_2$) nanocomposite was prepared via solid-state microwave irradiation by Yang et al.[93] and build asymmetric supercapacitor using $Fe_2O_3$-graphene and $MnO_2$-graphene as electrodes. The resulted supercapacitor displayed large potential window of 1.8 V which can combine high energy and power densities along with a good cycling life. Darvishi et al.[94] prepared graphene-CuO nanocomposite and studied the photocatalytic performance of the material which indicated that increase in graphene amount increases the average rate constant. Nanocomposite of rGO-titanium oxide was prepared by Ates et al.[95] using microwave irradiation. Supercapacitors made using the resulted nanocomposite displayed large specific capacitance with a high energy and power density which made it a suitable material for supercapacitor application. Nagaraju et al.[96] synthesized graphene-$WO_3$ nanocomposite using in situ microwave irradiation and the nanocomposite displayed specific capacitance of 761 $Fg^{-1}$ at the current density of 1 $Ag^{-1}$ with excellent cyclic stability indicates the potential use in supercapacitor.

---

[85] Fernando Langa et al., "Microwave Irradiation: More than Just a Method for Accelerating Reactions," *Contemp. Org. Synth.* 4 (1997): 373–86, doi:10.1039/CO9970400373.
[86] Ming Zhang et al., "Magnetite/graphene Composites: Microwave Irradiation Synthesis and Enhanced Cycling and Rate Performances for Lithium Ion Batteries," *Journal of Materials Chemistry* 20, no. 26 (2010): 5538–43, doi:10.1039/c0jm00638f.
[87] Jun Yan et al., "Fast and Reversible Surface Redox Reaction of Graphene-MnO2 Composites as Supercapacitor Electrodes," *Carbon* 48, no. 13 (2010): 3825–33, doi:10.1016/j.carbon.2010.06.047.
[88] Jun Yan et al., "Rapid Microwave-Assisted Synthesis of Graphene nanosheet/Co 3O4 Composite for Supercapacitors," *Electrochimica Acta* 55, no. 23 (2010): 6973–78, doi:10.1016/j.electacta.2010.06.081.
[89] K. S. Subrahmanyam et al., "A Study of Graphene Decorated with Metal Nanoparticles," *Chemical Physics Letters* 497, no. 1–3 (2010): 70–75, doi:10.1016/j.cplett.2010.07.091.
[90] Dorothea Marquardt et al., "The Use of Microwave Irradiation for the Easy Synthesis of Graphene-Supported Transition Metal Nanoparticles in Ionic Liquids," *Carbon* 49, no. 4 (2011): 1326–32, doi:10.1016/j.carbon.2010.09.066.

---

[91] Xianjun Zhu et al., "Nanostructured Reduced Graphene Oxide/Fe2O3 Composite As a High-Performance Anode Material for Lithium Ion Batteries," *ACS Nano* 5, no. 4 (2011): 3333–38, doi:10.1021/nn200493r.

[92] Kai Chih Hsu and Dong Hwang Chen, "Microwave-Assisted Green Synthesis of Ag/reduced Graphene Oxide Nanocomposite as a Surface-Enhanced Raman Scattering Substrate with High Uniformity," *Nanoscale Research Letters* 9 (2014): 193 (1-9), doi:10.1186/1556-276X-9-193.

[93] Minho Yang et al., "Three-Dimensional Expanded Graphene-Metal Oxide Film via Solid-State Microwave Irradiation for Aqueous Asymmetric Supercapacitors," *ACS Applied Materials and Interfaces* 7, no. 40 (2015): 22364–71, doi:10.1021/acsami.5b06187.

[94] Motahareh Darvishi, Ghafar Mohseni-Asgerani, and Jamileh Seyed-Yazdi, "Simple Microwave Irradiation Procedure for the Synthesis of CuO/Graphene Hybrid Composite with Significant Photocatalytic Enhancement," *Surfaces and Interfaces* 7, no. February (2017): 69–73, doi:10.1016/j.surfin.2017.02.007.

[95] Murat Ates et al., "Reduced Graphene oxide/Titanium Oxide Nanocomposite Synthesis via Microwave-Assisted Method and Supercapacitor Behaviors," *Journal of Alloys and Compounds* 728 (2017): 541–51, doi:10.1016/j.jallcom.2017.08.298.

**3.6 Ball milling method**

In ball milling[97] or mechanical alloying process a powder mixture is placed in a closed vessel and thereafter undergoes high-energy collision from the balls. In this high energy process graphite can be ruptured to produce graphene sheets which can be further mixed with other materials to produce nanocomposite. He et al.[98] prepared graphene- $Al_2O_3$ composite powders by ball milling method for different milling times. They concluded that addition of graphene hinders the grain growth of $Al_2O_3$ resulting in production of finer particles. Composites of GPL and powdered Al were prepared by Bartolucci et al.[99] using ball milling and compared with pure Al and multi-walled carbon nanotube composites. The comparative study reflected that strength and hardness were decreased for graphene-Al composite as a result of aluminium carbide formation. Bastwros et al.[100] prepared a 1.0 wt.% graphene reinforced aluminium (Al6061) composite via ball milling and reported that flexural strength was enhanced by 47% compared with the reference Al6061 processed at the same condition. GNP/Al composites were produced employing mechanical alloying by Bustamante et al.[101] and studied their mechanical behaviour which exhibited that hardness increases with milling time. Dutkiewicz et al.[102] used GPL and mixed it with Cu nanoparticles via milling process and the resulted nanocomposite revealed nearly 50% increase in hardness and about 30% decrease in electrical resistivity compared to the composite with coarse GPL. Tribological and mechanical properties of multilayer graphene reinforced $Ni_3Al$ matrix composites prepared by Zhai et al.[103] via ball milling. It was found that the elastic modulus and hardness of the metal nanocomposite were improved with increasing multilayer graphene content up to 1.0 wt.%, while decreased afterwards. Moreover addition of multilayer graphene decreases the friction coefficient and improves the wear resistance of the nanocomposite. Yue et al.[104] prepared GNS-Cu nanocomposite via ball milling method and studied the effect of variation of GNS content on elongation to fracture and ultimate tensile strength. The measurement revealed that GNS content in the composite must be 0.5 wt.% to achieve the best results. Du et al.[105] fabricated graphene-Mg nanocomposite using ball-milling of a pure Mg powder and GNS. Si@SiOx/graphene nanocomposites were prepared by Tie et al.[106] using ball milling technique which depicted high reversible capacity, enhanced cycling stability, and good rate capability to be used as high-performance anode materials for LIBs. Zhang et al.[107] prepared GNS-Al nanocomposite using ball milling which revealed increase in hardness by 115.1% due to the refined microstructure and Orowan strengthening.

---

[96] Perumal Nagaraju et al., "Rapid Synthesis of WO3/graphene Nanocomposite via in-Situ Microwave Method with Improved Electrochemical Properties," *Journal of Physics and Chemistry of Solids* 120 (2018): 250–60, doi:10.1016/j.jpcs.2018.04.046.

[97] Francesco Delogu, Giuliana Gorrasi, and Andrea Sorrentino, "Fabrication of Polymer Nanocomposites via Ball Milling: Present Status and Future Perspectives," *Progress in Materials Science* 86 (2017): 75–126, doi:10.1016/j.pmatsci.2017.01.003.

[98] Ting He et al., "Preparation and Consolidation of Alumina/Graphene Composite Powders," *Materials Transactions* 50, no. 4 (2009): 749–51, doi:10.2320/matertrans.MRA2008458.

[99] Stephen F. Bartolucci et al., "Graphene-Aluminum Nanocomposites," *Materials Science and Engineering A* 528, no. 27 (2011): 7933–37, doi:10.1016/j.msea.2011.07.043.

[100] Mina Bastwros et al., "Effect of Ball Milling on Graphene Reinforced Al6061 Composite Fabricated by Semi-Solid Sintering," *Composites Part B: Engineering* 60 (2014): 111–18, doi:10.1016/j.compositesb.2013.12.043.

---

[101] R. Pérez-Bustamante et al., "Microstructural and Hardness Behavior of Graphene-Nanoplatelets/aluminum Composites Synthesized by Mechanical Alloying," *Journal of Alloys and Compounds* 615, no. S1 (2015): S578–82, doi:10.1016/j.jallcom.2014.01.225.

[102] Jan Dutkiewicz et al., "Microstructure and Properties of Bulk Copper Matrix Composites Strengthened with Various Kinds of Graphene Nanoplatelets," *Materials Science and Engineering A* 628 (2015): 124–34, doi:10.1016/j.msea.2015.01.018.

[103] Wenzheng Zhai et al., "Investigation of Mechanical and Tribological Behaviors of Multilayer Graphene Reinforced Ni3Al Matrix Composites," *Composites Part B: Engineering* 70 (2015): 149–55, doi:10.1016/j.compositesb.2014.10.052.

[104] Hongyan Yue et al., "Effect of Ball-Milling and Graphene Contents on the Mechanical Properties and Fracture Mechanisms of Graphene Nanosheets Reinforced Copper Matrix Composites," *Journal of Alloys and Compounds* 691 (2016): 755–62, doi:10.1016/j.jallcom.2016.08.303.

[105] Xiaoming Du, Kaifeng Zheng, and Fengguo Liu, "Synthesis and Characterization of Graphene Nanosheets/magnesium Composites Processed through Powder Metallurgy," *MATERIALS AND TECHNOLOGY* 51, no. 6 (2017): 967–71, doi:10.17222/mit.2017.041.

[106] Xiaoyong Tie et al., "Si@SiOx/Graphene Nanosheets Composite: Ball Milling Synthesis and Enhanced Lithium Storage Performance," *Frontiers in Materials* 4, no. January (2018): 1–7, doi:10.3389/fmats.2017.00047.

## 3.7 Electrochemical deposition method

Electrochemical[108] deposition is synthesis method where a film of solid metal is deposited from a solution of ions onto an electrically conducting surface. This inexpensive and scalable method offers several advantages like high purity of deposited materials and rigid control of the composition. Yin et al.[109] deposited monocrystalline ZnO nanorods having high donor concentration on highly conductive rGO films and the resultant nanocomposite was incorporated in an inorganic–organic hybrid solar cell to increase its efficiency. Graphene-$MnO_2$ nanowall hybrids were synthesised by Zhu et al.[110] via a one-step electrochemical approach and the synthesised materials exhibited potential for applications in electrochemical biosensors and supercapacitors. Gao et al.[111] devised a simple single-step ultrasonication assisted electrochemical route for the synthesis of graphene-PtNi nanocomposites which showed excellent electrochemical performance in detecting nonenzymatic amperometric glucose in human urine samples. Jagannadham studied the thermal[112] and electrical[113] conductivity of the graphene-Cu nanocomposite films synthesized by electrochemical deposition and the results indicated that the resultant material is suitable for electrofriction applications. Graphene-Ni nanocomposites were synthesised employing electrodeposition in a nickel sulfamate solution with GO sheets by Kuang et al.[114]. Measurements on the nanocomposite showed increased thermal conductivity and significantly improved hardness. Xie et al.[115] synthesised rGO-Cu nanocomposite films with one-step electrochemical reduction deposition method which exhibited high electroactivity thus can be used in electrical contact materials. Ren et al.[116] successfully synthesized Graphene-Ni nanocomposite via electrochemical deposition and further examination revealed that elastic modulus of 240 GPa with a hardness of 4.6 GPa can be obtained with an addition of graphene as low as 0.05 g L$^{-1}$. GO-TiO$_2$ nanotube composite was synthesized via one-step anodization by Ali et al.[117] and the nanocomposite revealed a high photocatalytic activity for organics degradation under visible light.

---

[107] Jiangshan Zhang et al., "Microstructure and Mechanical Properties of Aluminium-Graphene Composite Powders Produced by Mechanical Milling," *Mechanics of Advanced Materials and Modern Processes* 4, no. 1 (2018): 4 (1-9), doi:10.1186/s40759-018-0037-5.
[108] Randa Abdel-Karim, "Electrochemical Synthesis of Nanocomposites," *Electrodeposition of Composite Materials*, 2016, 1–26, doi:10.5772/62189.
[109] Zongyou Yin et al., "Electrochemical Deposition of ZnO Nanorods on Transparent Reduced Graphene Oxide Electrodes for Hybrid Solar Cells," *Small* 6, no. 2 (2010): 307–12, doi:10.1002/smll.200901968.
[110] Chengzhou Zhu et al., "One-Step Electrochemical Approach to the Synthesis of Graphene/MnO2nanowall Hybrids," *Nano Research* 4, no. 7 (2011): 648–57, doi:10.1007/s12274-011-0120-2.
[111] H C Gao et al., "One-Step Electrochemical Synthesis of PtNi Nanoparticle-Graphene Nanocomposites for Nonenzynnatic Amperometric Glucose Detection," *ACS Applied Materials & Interfaces* 3 (2011): 3049–57, doi:10.1021/am200563f.
[112] K. Jagannadham, "Thermal Conductivity of Copper-Graphene Composite Films Synthesized by Electrochemical Deposition with Exfoliated Graphene Platelets," *Metallurgical and Materials Transactions B: Process Metallurgy and Materials Processing Science* 43, no. 2 (2012): 316–24, doi:10.1007/s11663-011-9597-z.
[113] K. Jagannadham, "Electrical Conductivity of Copper–graphene Composite Films Synthesized by Electrochemical Deposition with Exfoliated Graphene Platelets," *Journal of Vacuum Science & Technology B* 30, no. 3 (2012): 03D109 (1-9), doi:10.1116/1.3701701.

## 3.8 Other methods

Other than the common methods described before some other methods or mixed methods are also used for the synthesis of graphene-metal/ metal oxide nanocomposite. Sun et al.[118] employed heterogeneous coagulation method for the synthesis of graphene-TiO$_2$ nanocomposites and upon incorporation it enhanced the power conversion efficiency of dye-sensitized solar cells. Biju[119] used both hydrothermal and solution mixing method for the preparation of α-Fe$_2$O$_3$-graphene nanocomposite.

---

[114] Da Kuang et al., "Graphene-Nickel Composites," *Applied Surface Science* 273 (2013): 484–90, doi:10.1016/j.apsusc.2013.02.066.
[115] Guoxin Xie, Mattias Forslund, and Jinshan Pan, "Direct Electrochemical Synthesis of Reduced Graphene Oxide (rGO)/copper Composite Films and Their Electrical/electroactive Properties," *ACS Applied Materials and Interfaces* 6, no. 10 (2014): 7444–55, doi:10.1021/am500768g.
[116] Zhaodi Ren et al., "Mechanical Properties of Nickel-Graphene Composites Synthesized by Electrochemical Deposition," *Nanotechnology* 26, no. 6 (2015): 65706 (1-8), doi:10.1088/0957-4484/26/6/065706.
[117] I Ali et al., "One-Step Electrochemical Synthesis of Graphene Oxide-TiO2 Nanotubes for Improved Visible Light Activity," *Optical Materials Express* 7, no. 5 (2017): 1535–46, doi:10.1364/OME.7.001535.
[118] Shengrui Sun, Lian Gao, and Yangqiao Liu, "Enhanced Dye-Sensitized Solar Cell Using Graphene- TiO2 Photoanode Prepared by Heterogeneous Coagulation," *Applied Physics Letters* 96 (2010): 083113 (1-3), doi:10.1063/1.3318466.

## 4. Summary and outlook

Graphene-based metal nanocomposites demonstrated peerless potential which is reflected in their exponential increase in usage in various technological fields. As the requirements of high-performance graphene-metal/metal oxide nanocomposite increase with time, so much more research work will be carried out in this field. There are already seven major synthesis routes namely solution mixing, sol-gel, hydrothermal/solvothermal, self-assembly, microwave irradiation, ball milling, electrochemical deposition for the production of graphene-metal/metal oxide nanocomposite. The measurement of the resultant material depicted a great enhancement in mechanical, thermal and electrical properties. However, there are few key challenges remain related to synthesis methods, where scalable economical sustainable high quality graphene growth and reproducible homogeneous dispersion of graphene in metal matrix are most significant. The scalable economical[120] sustainable[121] high quality graphene growth is a matter of continuous challenge and researcher are seamlessly striving to address these issues. Functionalization of graphene[122] is proposed as a remedy to address the issues as it can modify the graphene metal interface in order to achieve homogeneous dispersion of graphene. Low bulk density of the graphene is another barrier to achieve homogeneous dispersion. All these studies indicate that the possibility of fabricating advanced graphene-based metal nanocomposites is strongly dependent on the graphene-metal interface engineering[123].

---

[119] C. S. Biju, "Properties of α-Fe2O3/graphene Nanohydrid Synthesized by a Simple Hydrothermal/solution Mixing Method," *Nano-Structures and Nano-Objects* 13 (2018): 44–50, doi:10.1016/j.nanoso.2017.12.005.
[120] Lixue Zou et al., "Trends Analysis of Graphene Research and Development," *Journal of Data and Information Science* 3, no. 1 (2018): 82–100, doi:10.2478/jdis-2018-0005.
[121] Michael Cai Wang et al., "A Sustainable Approach to Large Area Transfer of Graphene and Recycling of the Copper Substrate," *Journal of Materials Chemistry C* 5, no. 43 (2017): 11226–32, doi:10.1039/c7tc02487h.
[122] Xuezhong Gong et al., "Functionalized-Graphene Composites : Fabrication and Applications in Sustainable Energy and Environment," *Chemistry of Materials* 28 (2016): 8082−8118, doi:10.1021/acs.chemmater.6b01447.

---

[123] Xiang Zhang et al., "Effect of Interface Structure on the Mechanical Properties of Graphene Nanosheets Reinforced Copper Matrix Composites," *ACS Applied Materials and Interfaces* 10, no. 43 (2018): 37586–601, doi:10.1021/acsami.8b09799.

23